\documentclass[12pt,aps,manuscript,prb]{revtex4-1}
\usepackage[latin9]{inputenc}
\usepackage{geometry}
\usepackage{bm}
\usepackage{graphicx}
\usepackage{esint}

\makeatletter

\@ifundefined{textcolor}{}
{%
 \definecolor{BLACK}{gray}{0}
 \definecolor{WHITE}{gray}{1}
 \definecolor{RED}{rgb}{1,0,0}
 \definecolor{GREEN}{rgb}{0,1,0}
 \definecolor{BLUE}{rgb}{0,0,1}
 \definecolor{CYAN}{cmyk}{1,0,0,0}
 \definecolor{MAGENTA}{cmyk}{0,1,0,0}
 \definecolor{YELLOW}{cmyk}{0,0,1,0}
}

\makeatother

\begin{document}

\title{Revisiting the plasma sheath --- dust in plasma sheath}

\author{G. C. Das$^{1}$, R. Deka$^{2}$, and M. P. Bora$^{3}$}

\email{mpbora@gauhati.ac.in}

\affiliation{$^{1}$\emph{Mathematical Science Division, IASST, Guwahati 781014,
India.}\\
\emph{$^{2}$Physics Department, Gauhati University, Guwahati 781014,
India.}}
\begin{abstract}
In this work, we have considered the formation of warm plasma sheath
in the vicinity of a wall in a plasma with considerable presence of
dust particles. As an example, we have used the parameters relevant
in case of lunar plasma sheath, though the results obtained in this
work can be applied to any other physical situation such as laboratory
plasma. In the ion-acoustic time scale, we neglect the dust dynamics.
The dust particles affect the sheath dynamics by affecting the Poisson
equation which determines the plasma potential in the sheath region.
We have assumed the current to a dust particle to be balanced throughout
the analysis. This makes the grain potential dependent on plasma potential,
which is then incorporated into the Poisson equation. The resultant
numerical model becomes an initial value problem, which is described
by a 1-D integro-differential equation, which is then solved self-consistently
by incorporating the change in plasma potential caused by inclusion
of the dust potential in the Poisson equation.
\end{abstract}
\maketitle

\section{Introduction}

The physics of sheath in plasmas is an intriguing subject in fundamental
plasma physics and has been widely studied. Nevertheless, by no means
the study can be considered to be complete. The basic reason behind
this is the hugely varied categories of space and laboratory plasmas
with disparate physical conditions, where plasma sheaths form. These
phenomena range from basic laboratory plasmas (gas discharge devices)\cite{1,2,3,4}
to fusion plasma devices (viz.\ tokamaks)\cite{5} and from ionospheric
plasmas in planetary atmospheres\cite{6} to astrophysical plasmas
such as planetary nebulae\cite{7,8}. The plasma solitons can also
be considered as close cousins of stationary plasma sheaths, which
occur in laboratory as well as space plasmas and can be considered
as \emph{moving sheaths\cite{9}} in a plasma. Insight into sheath
dynamics requires the Poisson's equation to be solved with certain
assumptions for the ion and electron densities and only if the relevant
parameters are known. Tonks and Langmuir\cite{key-5} were among the
firsts to be able to reduce the complete sheath equation to a simpler
integral equation. Their solution gives the potential distribution
within the plasma. Auer\cite{key-6}, Caruso and Cavaliere\cite{key-7},
and Self\cite{key-8} have analyzed the phenomena to find out the
potential distribution solution throughout the plasma and in the sheath
region. It was Riemann\cite{16}, who presented a review regarding
the sheath formation and the basic features of sheath and its relation
to Bohm sheath criterion. As the study of plasma sheath has gained
attention due to its practical importance, researchers have been able
to establish a very good correlation between the theory and the experiment\cite{key-13,key-12,key-11,key-10}.

Though by plasmas we usually mean electron-ion plasmas, practically
they are never so pure and always have contaminations in the form
of dust. In extreme cases, these become dusty plasmas. In almost all
plasma sheaths, which usually form due to plasma-wall interactions,
some kind of impurities do intrude into the sheath. We in this work,
have studied such a phenomena which we refer to as \emph{dust in plasma
sheath}. As we know, dust particles are abundant in all kinds of plasmas
and they modify the plasma dynamics by being constituent plasma component
which happens as electrons (usually) settle down on the dust surfaces
due to their large electrical capacitance. In most cases, the charging
of the dust particles is treated in collisionless plasma physics,
theoretically with the well known Orbit Motion Limited (OML) theory\cite{10,11,12}.
As the dust particles are considerably heavier $(m_{d}\sim10^{15}m_{i})$,
their dynamical time scale is quite larger than the slowest electron-ion
time scale i.e.\ ion-acoustic time scale. However they do affect
the ion-acoustic dynamics by creating a charge imbalance in the electron-ion
population. Usually this is through electron depletion, although one
can fabricate different physical scenarios in laboratory plasmas\cite{13},
where positively charged dust particles are present.

Among different space plasma environments, the Moon provides a convenient
natural laboratory to study plasma-wall interactions in the space
environment. From various satellite observations, we now know that
a thin layer of dust $(\sim30\,{\rm cm})$ exists around the lunar
surface\cite{key-15}. Lunar Prospector (LP) and Apollo-era missions
gave the first hint that there must be a complex and coupled system
of plasmas and dust over the Moon's surface\cite{key-17}. The ambient
plasma and solar ultraviolet (UV) radiation incident on the lunar
surface are the main causes that the surface of the Moon becomes electrically
charged. On the dayside, photoemission typically dominates and the
lunar surface charges to a positive potential, while on the nightside
the plasma electrons usually dominate and the surface of the Moon
charges to a negative potential\cite{key-17,key-19}. As the lunar
surface (and other bodies without any protective environments) is
exposed to solar UV radiation, cosmic rays, and constant bombardment
by interplanetary bodies, a plasma sheath forms immediately above
the lunar surface into which dust particles are injected from the
regolith. This provides a very natural environment to study the complex
plasma sheath dynamics.

In this work, we have revisited the phenomena of plasma sheath with
a considerable presence of dust particles. In the ion-acoustic time-scale,
the the dust dynamics can be neglected and the dust particles are
considered to be stationary. The dust particles affect the sheath
dynamics by affecting the plasma potential. We have self-consistently
considered the dust-effects in the Poisson equation, which is then
solved to obtain the plasma and dust potentials. As we consider a
stationary sheath, we consider the current to the dust particles to
be balanced throughout the formation of the plasma sheath. As a result,
different currents to the surface of a dust particle become dependent
on the grain potential, which is a function of the particle size.
The resultant steady-state model is a 4-dimensional differential model,
which we have reduced to a one-dimensional integro-differential model,
owing to the integrability of the ion continuity and ion momentum
equations. Our results show that the dust-effect in the Poisson equation
considerably changes the structure of the dust layer, which might
form as a result of levitation due to oppositely directed electrostatic
and gravitational forces. As a probable application, we have used
parameters for lunar plasma environment, though the results obtained
in this work could be used in other plasma-wall interactions as well,
with appropriate parameters. In Sec.II, we consider our plasma model.
In Sec.III, we develop the sheath model. In Sec.IV, we consider the
dust-effects on the plasma sheath. In Sec.V, we consider contributions
of various forces on a dust particle inside a plasma sheath. Finally,
in Sec.VI, we summarise our results.

\section{Plasma model}

Our basic equations are due to a collisionless electron-ion plasma
with a considerable presence of dust grains. The equations (in 1-D)
are ion continuity and momentum equations and electrons are assumed
to be Boltzmannian owing to their negligible mass, 
\begin{eqnarray}
\frac{\partial n_{i}}{\partial t}+\frac{\partial}{\partial x}(n_{i}u_{i}) & = & 0,\\
\frac{\partial u_{i}}{\partial t}+u_{i}\frac{\partial u_{i}}{\partial x} & =- & \frac{1}{m_{i}n_{i}}\frac{\partial p_{i}}{\partial x}-\frac{e}{m_{i}}\frac{\partial\phi}{\partial x},\\
n_{e} & = & n_{0}e^{e\phi/T_{e}},
\end{eqnarray}
where $\phi$ is the plasma potential and the temperature is expressed
in energy unit. We use the ion equation of state as,
\begin{equation}
p_{i}\propto n_{i}^{\gamma},
\end{equation}
The model is closed by the Poisson equation,
\begin{equation}
\epsilon_{0}\frac{\partial^{2}\phi}{\partial x^{2}}=e(n_{e}-n_{i}+z_{d}n_{d}).
\end{equation}
All the symbols have their usual meanings. Note that the presence
of dust grains are incorporated into the model through the Poisson
equation. The number of charge that resides on the surface of a dust
grain is a function of the grain potential $\phi_{g}$. We have chosen
to normalize the plasma potential with $T_{e}/e$, length with Debye
length $\lambda_{D}$, velocity with ion-thermal velocity $u_{s}=\sqrt{T_{e}/m_{i}}$,
time with $\lambda_{D}/u_{s}$, electron, ion, and dust densities
with the respective equilibrium densities $n_{i0,e0}$, and ion pressure
$p_{i}$ with $n_{i0}T_{i}$. We also denote the ion to electron temperature
ratio by $\sigma=T_{i}/T_{e}$. The normalized equations now read
as, 
\begin{eqnarray}
\frac{\partial n_{i}}{\partial t}+\frac{\partial}{\partial x}(n_{i}u_{i}) & = & 0,\label{eq:cont}\\
\frac{\partial u_{i}}{\partial t}+u_{i}\frac{\partial u_{i}}{\partial x}+\frac{\sigma}{n_{i}}\frac{\partial p_{i}}{\partial x} & = & -\frac{\partial\phi}{\partial x},\label{eq:mom}\\
n_{e} & = & e^{\phi},\label{eq:ne}\\
p_{i} & = & n_{i}^{\gamma},\\
\frac{\partial^{2}\phi}{\partial x^{2}} & = & n_{e}-\delta_{i}n_{i}+\delta_{d}z_{d},\label{eq:poisson}
\end{eqnarray}
where $\delta_{i}=n_{i0}/n_{e0}$ and $\delta_{d}=z_{d0}n_{d0}/n_{e0}$
are the equilibrium density ratios.

\section{Sheath equations}

\begin{figure}
\begin{centering}
\includegraphics[width=0.5\textwidth]{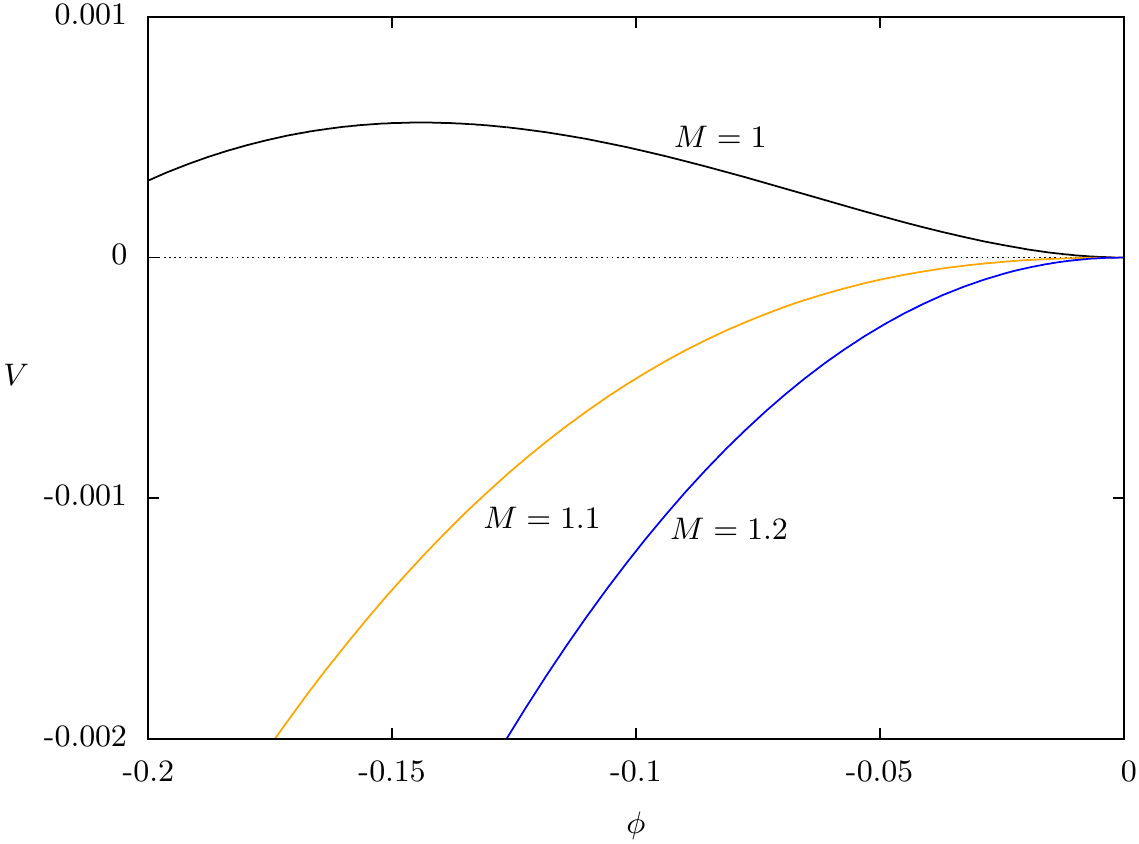}
\par\end{centering}

\protect\caption{\label{fig:pseudo-pot}The shape of the pseudo potential for different
Mach number for $\sigma=0.1$ when dust effects are neglected. Note
that we must have $M>1$ for $\sigma>0$ in a warm sheath with $\gamma=5/3$
for real solution of Eq.(\ref{eq:sheath equation}). The allowed region
is given by $V<0$.}
\end{figure}

Consider now a stationary plasma sheath\cite{13}. Far away from the
sheath, the plasma potential vanishes and other plasma parameters
approaches their bulk (equilibrium) values i.e.\ as $x\to\infty$,
$\phi\to0,u_{i}\to u_{0}\equiv M,p_{i}\to1,n_{i}\to1,z_{d}=z_{d}/z_{d0}\to1$,
$M$ is the Mach number and $z_{d0}=z_{d}(\phi)|_{\phi=0}$ is the
dust charge number in the bulk plasma. The steady state dynamical
equations are,
\begin{eqnarray}
\frac{\partial}{\partial x}(n_{i}u_{i}) & = & 0,\\
-u_{i}\frac{\partial u_{i}}{\partial x}+\frac{\sigma}{n_{i}}\frac{\partial p_{i}}{\partial x} & = & -\frac{\partial\phi}{\partial x}.\label{eq:cons}
\end{eqnarray}
From the continuity equation, we have
\begin{equation}
n_{i}=M/u_{i}.\label{eq:eqn}
\end{equation}
Integration of Eq.(\ref{eq:cons}) results the conservation of total
energy flux which is a combination of the kinetic flux, enthalpy flux,
and electrostatic flux,
\begin{equation}
\phi=\frac{1}{2n_{i}^{2}}M^{2}\left(n_{i}^{2}-1\right)+\frac{\gamma\sigma}{(\gamma-1)}\left(1-n_{i}^{\gamma-1}\right).\label{eq:phi}
\end{equation}
An expression for $n_{i}$ as a function of $\phi$ can be found from
Eqs.(\ref{eq:eqn},\ref{eq:phi}), which \emph{must} be solved numerically
for arbitrary $\gamma$. For $\gamma=3$ however, we can find an analytical
expression for $n_{i}(\phi)$ as,
\begin{equation}
n_{i}=\frac{1}{2\sqrt{3\sigma}}\left[\left\{ \left(M+\sqrt{3\sigma}\right)^{2}-2\phi\right\} ^{1/2}-\left\{ \left(M-\sqrt{3\sigma}\right)^{2}-2\phi\right\} ^{1/2}\right].\label{eq:ni}
\end{equation}
The signs in front of the square roots are fixed through the boundary
condition on $n_{i}$. Expressing ion density as a function of the
plasma potential, $n_{i}\equiv n_{i}(\phi)$, the Poisson's equation
can be integrated to get,
\begin{equation}
\frac{1}{2}\left(\frac{d\phi}{d\xi}\right)^{2}+V(\phi,M,\sigma,\gamma)=0,\label{eq:sheath equation}
\end{equation}
where $V(\phi,M,\sigma,\gamma)$ is the equivalent Sagdeev potential
or pseudo potential\cite{14} for a sheath, given by,
\begin{equation}
V(\phi,M,\sigma,\gamma)=1-e^{\phi}+\delta_{i}\int_{0}^{\phi}n_{i}(\phi)\, d\phi-\delta_{d}\int_{0}^{\phi}z_{d}(\phi)\, d\phi,\label{eq:pseduo-potential}
\end{equation}
where we have neglected the variation in the dust density keeping
in view that dust particles can be considered to be stationary in
the ion-acoustic time scale. For real solution, $V(\phi,M,\sigma,\gamma)<0$
for all values of $\phi$. This also determines the minimum velocity
for the ions $(u_{0}\equiv M)$ at the sheath boundary (the Bohm condition\cite{15,16}).
The boundary condition on $V$ is: at $\phi=0$, $V(\phi)=0$. In
Fig.\ref{fig:pseudo-pot}, the pseudo potential for various Mach numbers
are shown for $\phi<0$ with $\delta_{d}=0$ (refer to Sec.3.3) and
$\gamma=5/3$. 

\subsection{The modified Bohm condition}

The condition for ions to have a minimum velocity at the sheath boundary
for which $V(\phi,M,\sigma,\gamma)<0$, gives rise to the so called
Bohm condition\cite{15,16}. In general, for warm ions, we \emph{must}
find out the Bohm condition numerically. For warm ions with $\gamma=3$
and $\delta_{d}=0$, we can find out the Bohm condition by demanding
a local maxima for $V_{\gamma=3}(\phi,M,\sigma)|_{\phi\to-0}$,
\begin{equation}
M^{2}\geq1+3\sigma,
\end{equation}
which reduces to the classical Bohm condition, $M^{2}\geq1$ for $\sigma=0$\cite{4}.

\section{Dust in plasma sheath}

\subsection{Wall potential}

The electron distribution far away from the wall (or sheath) is given
by the usual Maxwell's distribution\cite{3,13},
\begin{equation}
f_{e}(v_{e})=n_{0}\left(\frac{m_{e}}{2\pi T_{e}}\right)^{3/2}\exp\left(-\frac{m_{e}v_{e}^{2}}{2T_{e}}\right).
\end{equation}
The electron current at the wall is due to those electrons which can
reach the wall with a minimum velocity $v_{{\rm min}}$ to overcome
the negative potential at the wall $\phi_{w}$\cite{13},
\begin{equation}
v_{{\rm min}}=\left(-\frac{2e\phi_{w}}{m_{e}}\right)^{1/2},
\end{equation}
from which we get the electron current density at the wall as,
\begin{eqnarray}
j_{e} & = & -e\int_{v_{{\rm min}}}^{\infty}\int_{-\infty}^{\infty}\int_{-\infty}^{\infty}vf_{e}(v)\, d\bm{v},\nonumber \\
 & = & -n_{0}e\left(\frac{T_{e}}{2\pi m_{e}}\right)^{1/2}\exp\left(\frac{e\phi_{w}}{T_{e}}\right).
\end{eqnarray}
The ion current is given by\cite{13},
\begin{equation}
j_{i}=en_{i}u_{i}\left(\frac{T_{e}}{m_{i}}\right)^{1/2}\equiv en_{0}u_{0}\left(\frac{T_{e}}{m_{i}}\right)^{1/2},
\end{equation}
where $u_{0}$ is the ion velocity (dimensional) at the sheath boundary
(the Mach number). For a stationary sheath, we must have $j_{e}+j_{i}=0$,
which determines the wall potential (normalized)\cite{6},
\begin{equation}
\phi_{w}=-2.84+\ln M,\label{eq:wall potential}
\end{equation}
where $M$ must have value for which $V(\phi,M,\sigma,\gamma)<0$.

\subsection{Current balance}

\begin{figure}
\begin{centering}
\includegraphics[width=0.5\textwidth]{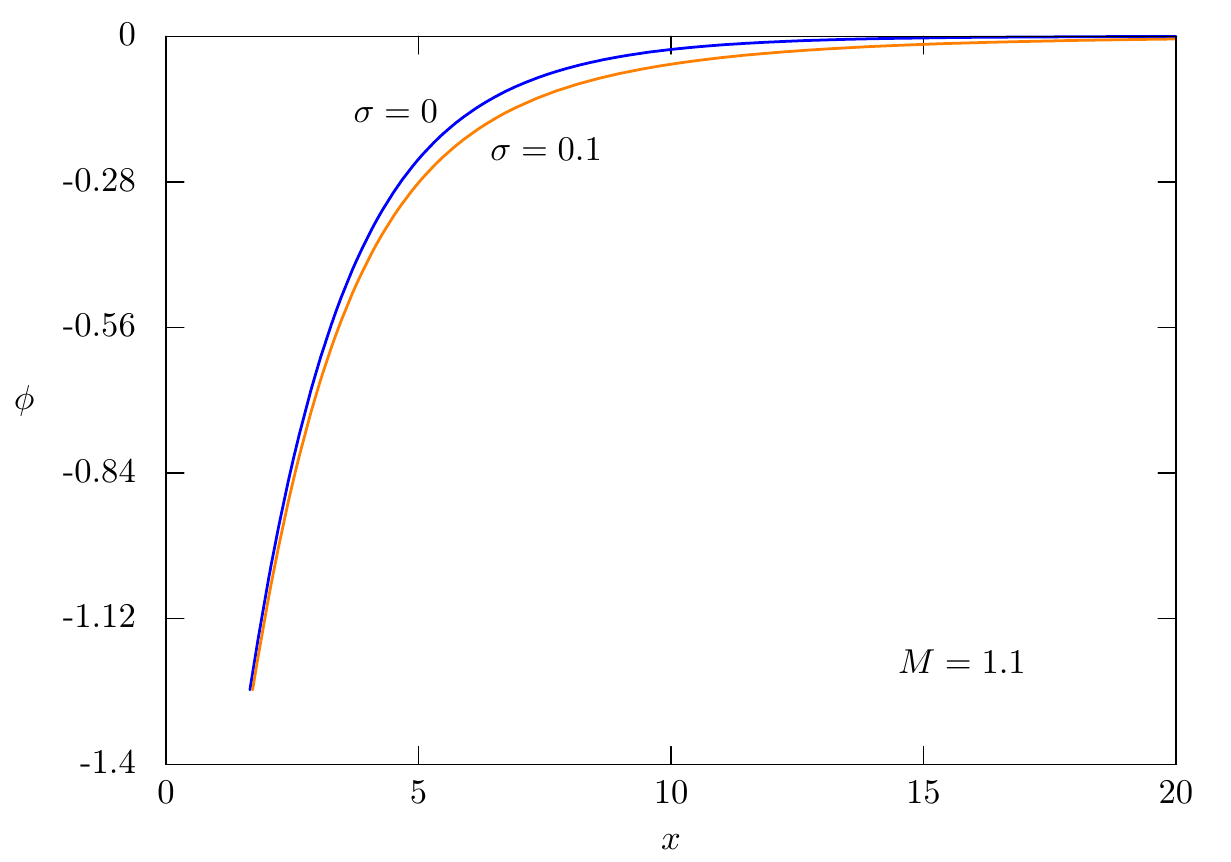}\hfill{}\includegraphics[width=0.5\textwidth]{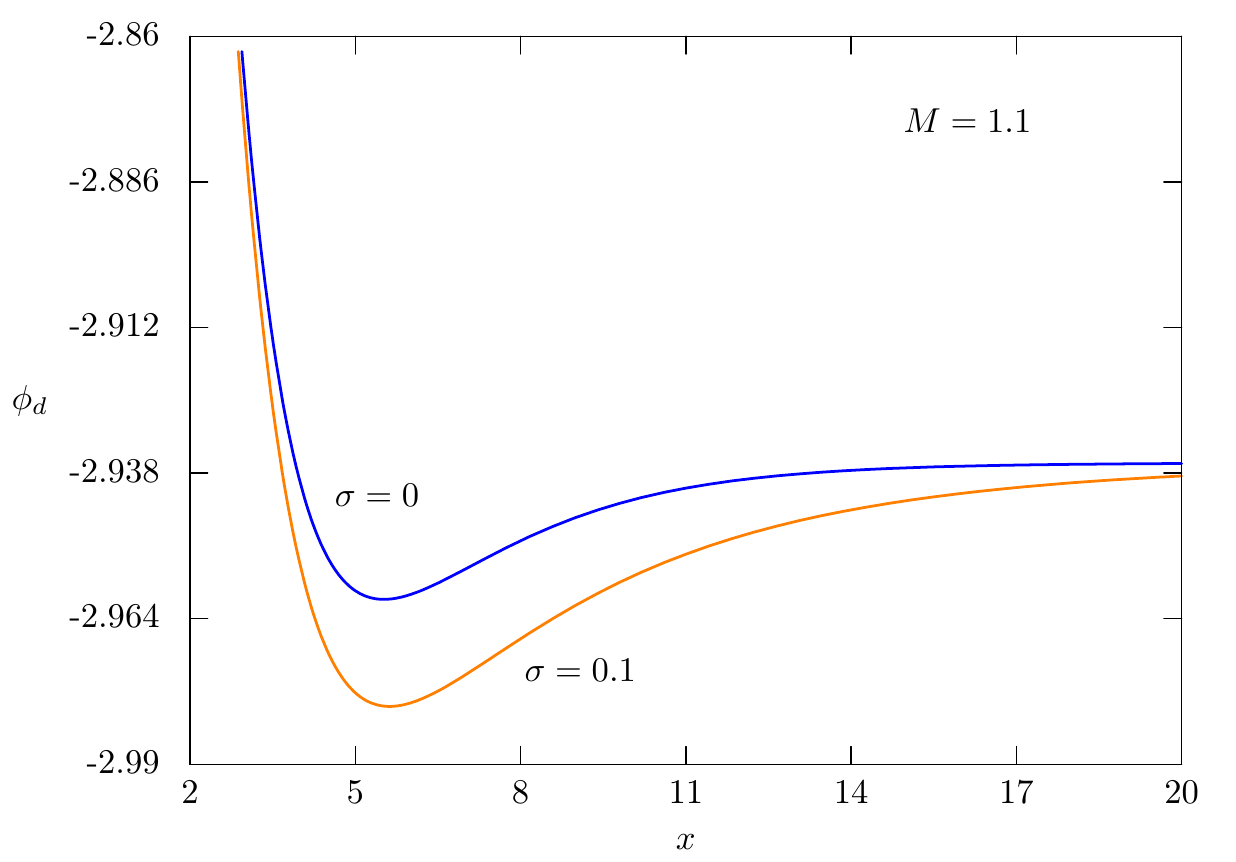}
\par\end{centering}

\protect\caption{\label{fig:potentials}The plasma potential $\phi$ and dust potential
$\phi_{d}$ as a function of the normalized distance $x$ for cold
$(\sigma=0)$ and warm $(\sigma>0)$ sheath with $\gamma=5/3$. The
wall is at $x=0$.}
\end{figure}

We assume that the current to the surface of dust particles remain
balanced \emph{all throughout} the plasma including the sheath\cite{13}.
So at any instant, we \emph{must} have,
\begin{equation}
I_{e}+I_{i}=0.\label{eq:balance}
\end{equation}
We further assume that the dust particles are negatively charged.
The ion current to the dust particles (dimensional) can be written
as\cite{13},
\begin{equation}
I_{i}=en_{i}u_{i}\sigma_{i},\label{eq:ion-current}
\end{equation}
where $\sigma_{i}=\pi b^{2}$ is the collision cross-section for ions
with the dust particles, $b$ being the impact parameter. Conservation
of angular momentum in an ion-dust collision leads to the condition,
\begin{equation}
u_{i}b=u_{g}r_{d},\label{eq:conservation-ug}
\end{equation}
where $u_{g}$ is the grazing velocity for the ions\cite{6,13} with
respect to a collision-event with dust particles and $r_{d}$ is the
radius of a dust particle. From energy consideration, we further have
\begin{equation}
\frac{1}{2}m_{i}u_{i}^{2}=\frac{1}{2}m_{i}u_{g}^{2}+e\phi_{d},\label{eq:energy-balance-collision}
\end{equation}
where $\phi_{d}=\phi_{g}-\phi$, $\phi_{g}$ being the grain potential.
Using Eqs.(\ref{eq:ion-current}-\ref{eq:energy-balance-collision})
we can write the normalized expression for $I_{i}$ as,
\begin{equation}
I_{i}=\pi r_{d}^{2}n_{i}u_{i}\left(1-\frac{2\phi_{d}}{u_{i}^{2}}\right),
\end{equation}
where $r_{d}$ is the dust radius measured in terms of Debye length.
Through Eq.(\ref{eq:eqn}), the above equation becomes
\begin{equation}
I_{i}=\pi r_{d}^{2}M\left(1-\frac{2\phi_{d}}{M^{2}}n_{i}^{2}\right)
\end{equation}
in which we need to substitute $n_{i}$ as a function of $\phi$ by
solving Eq.(\ref{eq:phi}). We note that for cold ions, the ion current
becomes
\begin{equation}
I_{i}^{{\rm cold}}=\pi r_{d}^{2}M\left(1-\frac{2\phi_{d}}{M^{2}-2\phi}\right),
\end{equation}
which can be obtained by substituting $\sigma=0$ in Eq.(\ref{eq:phi}).
For negatively charged dust particles, the normalized electron current
takes the form\cite{13},
\begin{equation}
I_{e}=-\sqrt{8}\pi r_{d}^{2}\delta_{m}e^{\phi+\phi_{d}},
\end{equation}
where $\delta_{m}=\sqrt{m_{i}/m_{e}}\approx43$. The current balance
equation, Eq.(\ref{eq:balance}) is now given by,
\begin{equation}
M\left(1-\frac{2\phi_{d}}{M^{2}}n_{i}^{2}\right)-\sqrt{8}\delta_{m}e^{\phi+\phi_{d}}=0,\label{eq:normalised current balance}
\end{equation}
Note that for spherical dust particles, the total amount of charge
$Q_{d}$ on the surface of a dust particles can be written as,
\begin{equation}
Q_{d}=C\,\Delta V=4\pi\epsilon_{0}r_{d}\phi_{d},\label{eq:qd}
\end{equation}
where $C$ is the grain capacitance. The dust potential $\phi_{d}$
is the principal solution of Eq.(\ref{eq:normalised current balance})
and is shown in Fig.\ref{fig:potentials} (right panel) against the
normalized distance $x$ for cold and warm sheath. The finite ion
temperature $(\sigma>0)$ affects the dust potential through the ion
density $n_{i}$. In this figure, the plasma potential $\phi$ is
a solution of the Poisson equation with $\delta_{d}=0$. Warm ions
causes the sheath to expand much like a Debye shielding expanding
with temperature. This can be seen from the behaviour of the potentials
for warm sheaths, which approaches the bulk value at a larger $x$
than in cold sheath.

\subsubsection{Other contributions to $I_{e,i}$}

We note that apart from the electron and ion currents to the dust
particles, there are other sources\cite{17,18}. For example in case
of dust particles near lunar surface, photoelectron current $I_{{\rm ph}}$
emitted by the lunar surface and net photoemission current $I_{h\nu}$
emitted by the dust particles may become significant\cite{17}. For
$\phi_{d}<0$, which usually is the case, we have
\begin{eqnarray*}
I_{{\rm ph}} & = & \frac{1}{4}J_{{\rm ph}}\pi r_{d}^{2},\\
I_{h\nu} & = & J_{h\nu}\pi r_{d}^{2}\exp\left(\frac{e\phi_{d}}{T_{{\rm ph}}}\right),
\end{eqnarray*}
where
\[
J_{h\nu}=-en_{{\rm ph}}\left(\frac{T_{{\rm ph}}}{2\pi m_{e}}\right),
\]
is the photoemission current density and $J_{{\rm ph}}\sim4.5\,\mu{\rm A}/{\rm m}$
is the photoelectron current density. However, in this work for the
sake of simplicity we have omitted these currents as contributions
from both these currents can be incorporated into the ion and electron
currents. 

\subsection{Negligible dust}

When number of dust particles are considerably less than that of ions
and electrons $(n_{d}\ll n_{i,e})$, $\delta_{d}$ can be neglected
and the pseudo potential given by Eq.(\ref{eq:pseduo-potential})
can be evaluated without any effect from the dust,
\begin{equation}
V(\phi,M,\sigma,\gamma)=1-e^{\phi}+\delta_{i}\int_{0}^{\phi}n_{i}(\phi)\, d\phi.
\end{equation}
In such case, for $\gamma=3$, we can analytically evaluate the above
integration in, 
\begin{eqnarray}
V_{\gamma=3}(\phi,M,\sigma,\delta_{d}=0) & = & \left(1-e^{\phi}\right)-\frac{1}{6\sqrt{3\sigma}}\left[\left\{ \left(M+\sqrt{3\sigma}\right)^{2}-2\phi\right\} ^{3/2}-\left(M+\sqrt{3\sigma}\right)^{3}\right.\label{eq:pot}\\
 &  & \left.-\,\left\{ \left(M-\sqrt{3\sigma}\right)^{2}-2\phi\right\} ^{3/2}+\left(M-\sqrt{3\sigma}\right)^{3}\right].\nonumber 
\end{eqnarray}
When the ion temperature is also neglected $(\sigma=0)$, we get the
classical sheath equation\cite{4},
\begin{equation}
V_{\sigma=0}(\phi,M,\delta_{d}=0)=M^{2}\left[\left(1-\frac{2\phi}{M^{2}}\right)^{1/2}-1\right]+e^{\phi}-1.\label{eq:pot-td0}
\end{equation}
The above expression can also be obtained by a series expansion of
Eq.(\ref{eq:pot}) or Eq.(\ref{eq:pseduo-potential}) in the limit
$\sigma\to0$ when $\delta_{d}=0$ (equivalently $\delta_{i}=1$)
.

\subsection{With dust effects}

When there are considerable presence of dust particles, they begin
to affect the pseudo potential through the Poisson equation and we
need to use the full form of the Poisson equation, Eq.(\ref{eq:poisson}).
This procedure requires numerical solution of the plasma model. From
Eqs.(\ref{eq:poisson},\ref{eq:ni},\ref{eq:normalised current balance}),
the complete numerical model can be summarised as,
\begin{eqnarray}
\frac{d^{2}\phi}{dx^{2}} & = & e^{\phi}-\delta_{i}{\cal N}(\phi)+\delta_{d}\frac{{\cal F}(\phi)}{{\cal F}(0)},\label{eq:npois}\\
2\phi{\cal N}(\phi)^{2} & = & M^{2}\left[{\cal N}(\phi)^{2}-1\right]+\frac{2\gamma\sigma}{(\gamma-1)}{\cal N}(\phi)^{2}\left[1-{\cal N}(\phi){}^{\gamma-1}\right],\label{eq:nneq}\\
\sqrt{8}\delta_{m}e^{\phi+{\cal F}(\phi)} & = & M\left[1-\frac{2}{M^{2}}{\cal F}(\phi)\,{\cal N}(\phi)^{2}\right],\label{eq:nbal}
\end{eqnarray}
where ${\cal N}(\phi)\equiv n_{i},{\cal F}(\phi)\equiv\phi_{d}$ are
\emph{numerical} functions of $\phi$ corresponding to the ion density
and dust potential. The function ${\cal N}(\phi)$ can be evaluated
by solving Eq.(\ref{eq:nneq}) which when substituted in Eq.(\ref{eq:nbal}),
can be solved for ${\cal F}(\phi)$. Poisson equation Eq.(\ref{eq:npois})
then, \emph{must} be solved self-consistently to evaluate $\phi(x)$,
which can be used to generate $\phi_{d}(x)$ and $n_{i}(x)$.

Note that the above system of equations can also be cast as a fully
differentiable system,
\begin{eqnarray}
\frac{d^{2}\phi}{dx^{2}} & = & e^{\phi}-\delta_{i}{\cal N}(\phi)+\delta_{d}\frac{{\cal F}(\phi)}{{\cal F}(0)},\label{eq:np}\\
\frac{d{\cal N}}{dx} & = & -\frac{{\cal N}^{3}}{M^{2}+\gamma\sigma{\cal N}^{\gamma+1}}\frac{d\phi}{dx},\label{eq:nn}\\
\frac{d{\cal F}}{dx} & = & -\frac{d\phi}{dx}+\frac{{\cal N}}{{\cal N}^{2}+\sqrt{2}\delta_{m}Me^{\phi+{\cal F}}}\left({\cal N}\frac{d\phi}{dx}-2{\cal F}\frac{d{\cal N}}{dx}\right).\label{eq:nb}
\end{eqnarray}
Many authors use the system as given by Eqs.(\ref{eq:np}-\ref{eq:nb})
to solve the sheath problem with some initial conditions, usually
starting from the bulk plasma\cite{19,20}. However, we note that
starting with initial conditions in the bulk plasma (i.e.\ starting
from the beginning of the sheath region and integrating toward the
wall) is not suitable in many cases as all the derivatives in the
above equations vanishes in the bulk plasma, thereby numerically rendering
it impossible to evolve. To circumvent this, the standard procedure
is to resort to a series expansion at the sheath boundary\cite{19,20},
which is essentially like \emph{pushing} the solution through its
desired track and one may often arrive at erroneous values of the
variables by the time the solution reaches the wall, depending on
accuracy of the series expansion and the position of the sheath boundary.
Besides, as Eqs.(\ref{eq:np}-\ref{eq:nb}) is a four-dimensional
system, in certain parameter regime the system may become highly sensitive
to the initial conditions\cite{21} and one may arrive at a completely
different solution from what it was intended to. In our opinion, the
differential system is best solved as a multi-dimensional boundary
value problem (BVP). However solving a multi-dimensional BVP has its
own demerits\cite{22}. 
\begin{figure}
\begin{centering}
\includegraphics[width=0.5\textwidth]{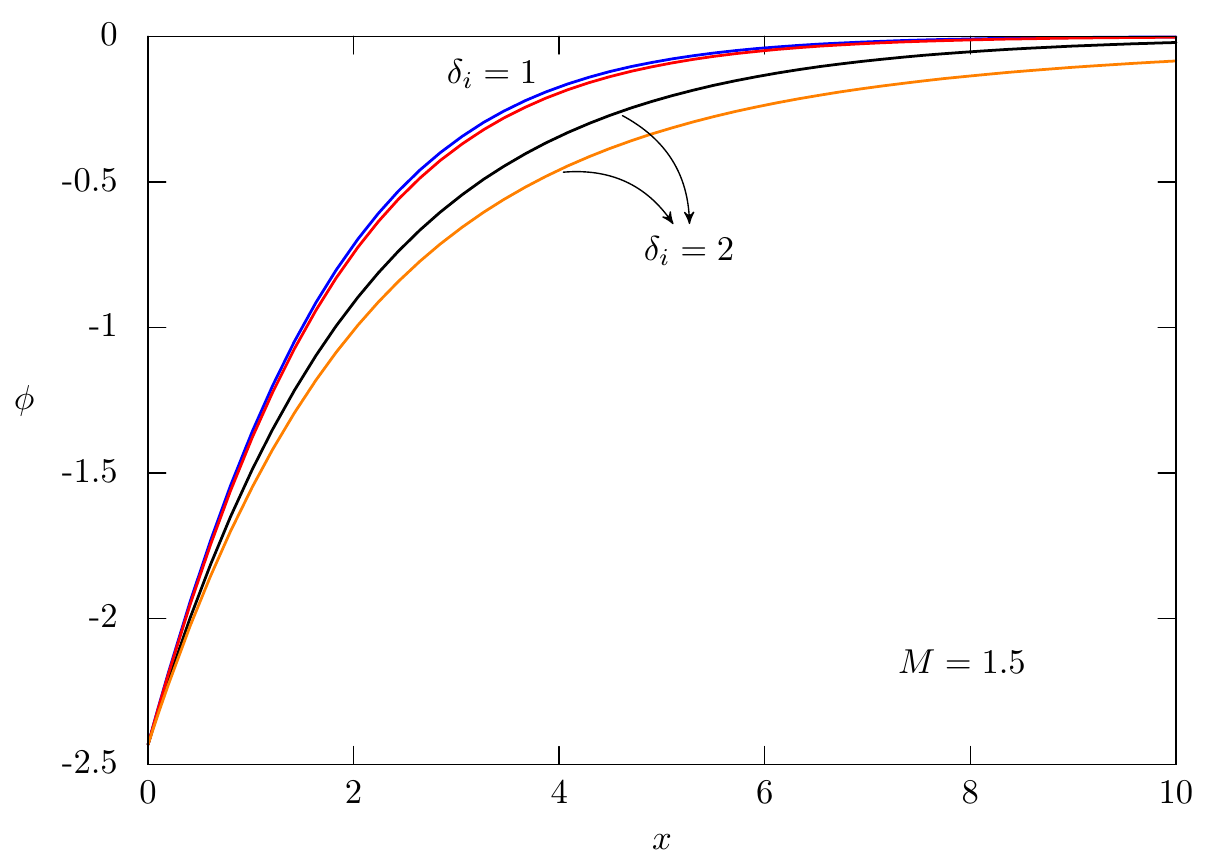}\hfill{}\includegraphics[width=0.5\textwidth]{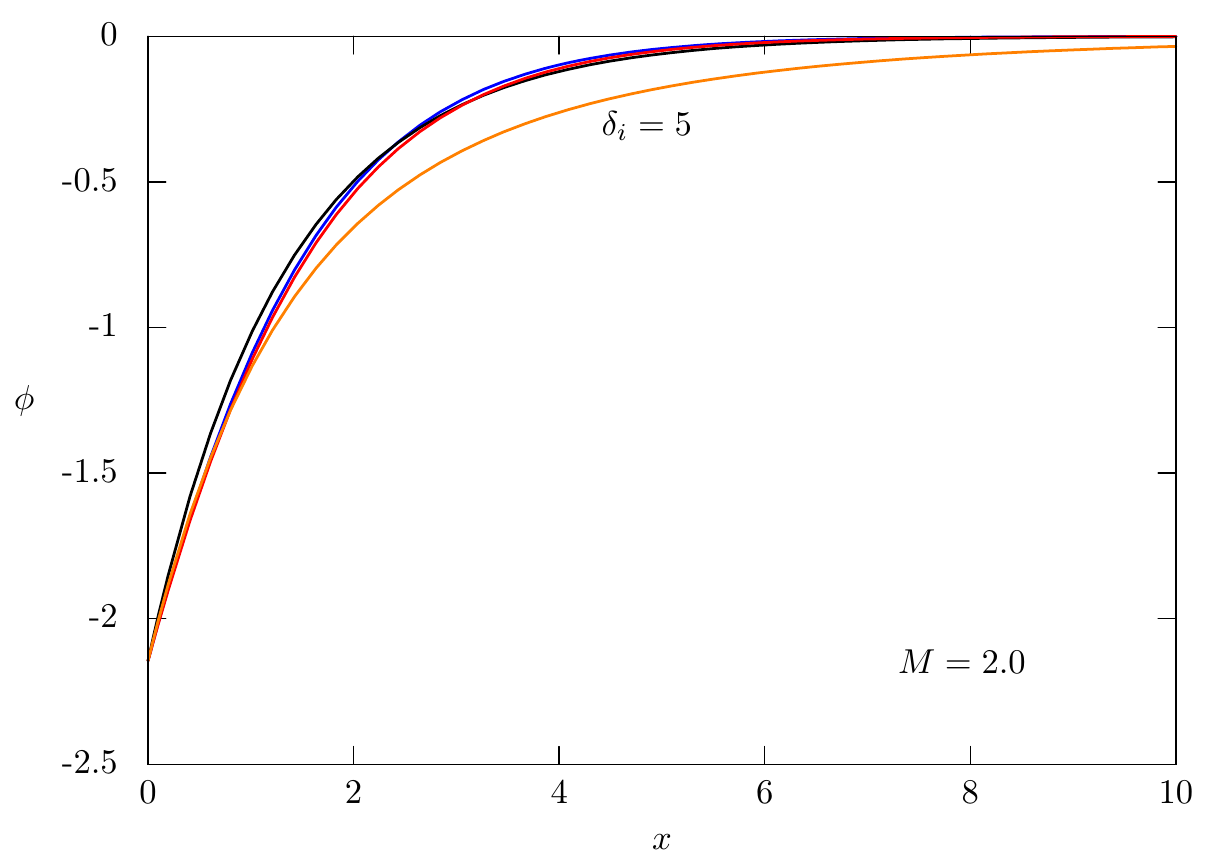}\\
\includegraphics[width=0.5\textwidth]{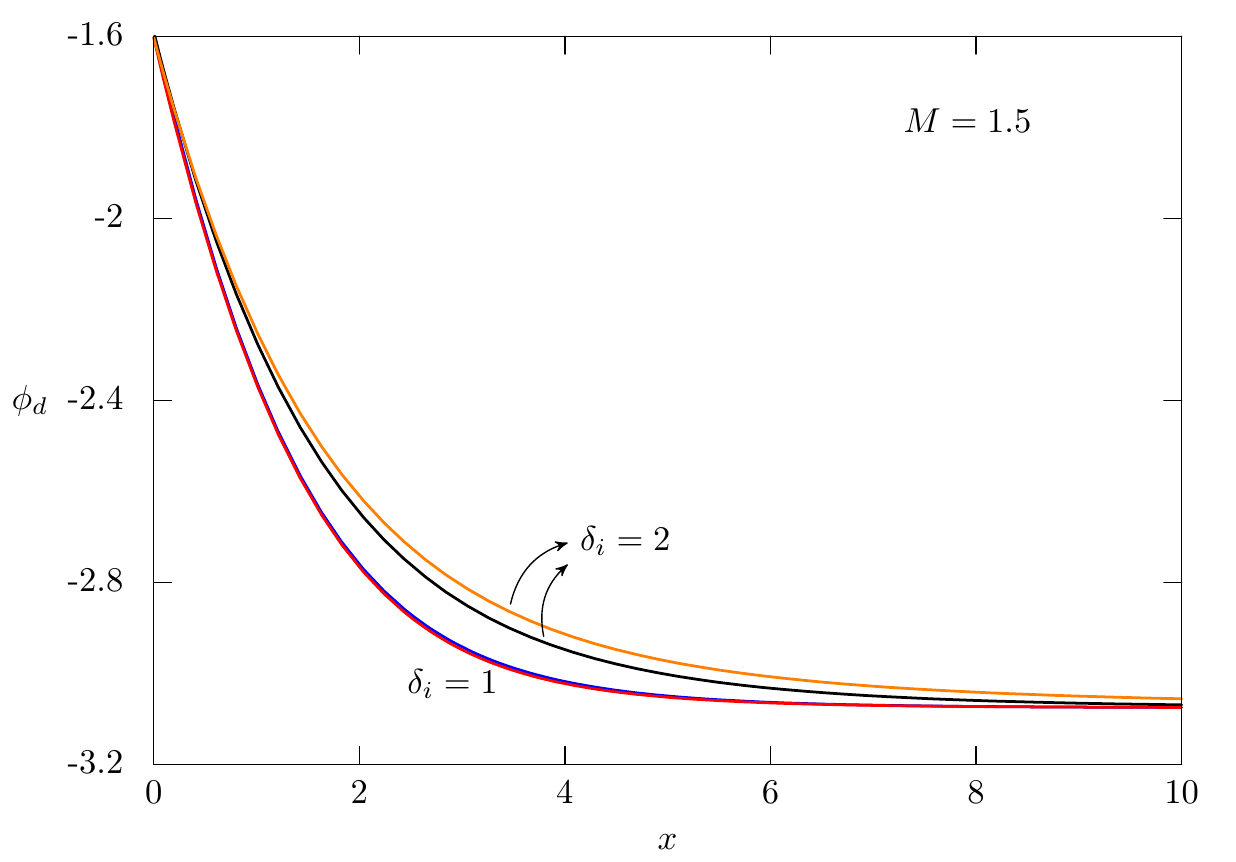}\hfill{}\includegraphics[width=0.5\textwidth]{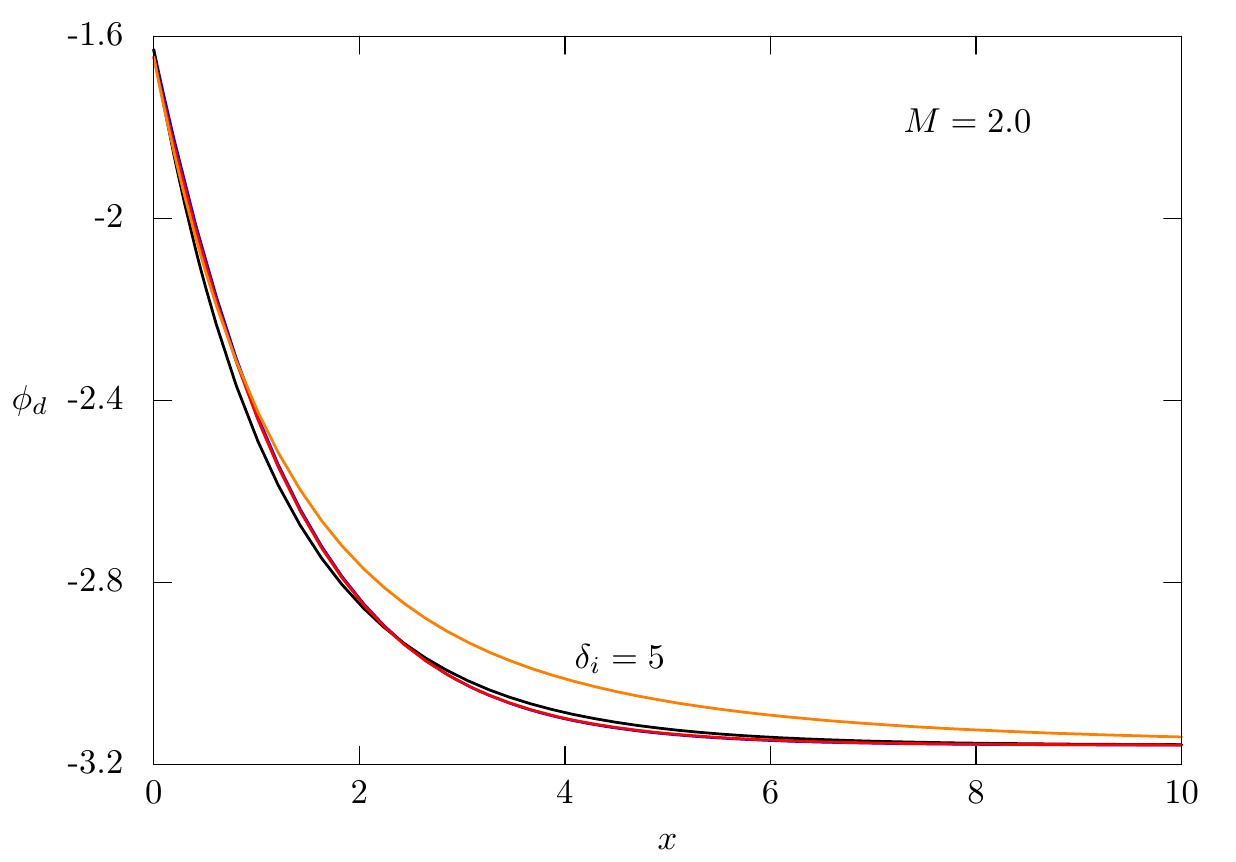}
\par\end{centering}

\protect\caption{\label{fig:potentials-1}The variation of plasma and dust potentials
with normalized distance from the wall $(x)$ with $(\delta_{i}>1)$
and without dust effects $(\delta_{i}=1)$, which are indicated in
the figures. The blue $(\sigma=0)$ and red $(\sigma>0)$ colored
curves are for $\delta_{i}=1$. The black $(\sigma=0)$ and orange
$(\sigma>0)$ colored curves are for $\delta_{i}>1$. The values of
$(M,\sigma,\delta_{i})$ for the left column are $M=1.5,\delta_{i}=(1,2),\sigma=(0,0.2)$
and the right column are for $M=2,\delta_{i}=(1,5),\sigma=(0,0.5)$. }
\end{figure}

In this work, we choose to solve the system as given by Eqs.(\ref{eq:npois}-\ref{eq:nbal})
by self consistently solving Eq.(\ref{eq:npois}) along with Eqs.(\ref{eq:nneq},\ref{eq:nbal}).
This is done by writing the Poisson equation in terms of the pseudo
potential as an integro-differential equation,
\begin{equation}
\frac{1}{\sqrt{2}}\frac{d\phi}{dx}=\left[e^{\phi}+(\delta_{i}-1)\int_{0}^{\phi}{\cal F}(\psi)\, d\psi-\delta_{i}\int_{0}^{\phi}{\cal N}(\psi)\, d\psi-1\right]^{1/2},
\end{equation}
where we have substituted $\delta_{d}=\delta_{i}-1$ (from the quasi-neutrality
condition). We have solved the above equation with the help of a modified
4th order Runge-Kutta method with adaptive step-size control\cite{22}
and evaluating the integrations numerically in each step. The differential
equation is solved from the wall $(x=0)$ toward the bulk plasma $(x\to\infty)$
with the single initial condition $\phi(x=0)=\phi_{w}$ as given by
Eq.(\ref{eq:wall potential}). The evaluation of the integrations
requires Eqs.(\ref{eq:nneq},\ref{eq:nbal}) to be solved self-consistently,
which is done with the help of Newton's method\cite{22}. As we have
the exact initial condition at our disposal, the solution can be built
very accurately into the bulk plasma, as far as required. Practically
we solve from $x=0$ (wall) to a large number, say $x\sim50$ (bulk
plasma).

The results are summarised in Fig.\ref{fig:potentials-1} where we
have shown the potentials for both cold $(\sigma=0)$ and warm $(\sigma>0)$
sheaths, with $(\delta_{i}>1)$ and without $(\delta_{i}=1)$ dust
effects. The results are shown for two sets of values, each for the
Mach number $M$, $\sigma$, and $\delta_{i}$ (for details, see figure
caption). Note that a value of $\delta_{i}=2$ indicates $50\%$ depletion
of the electrons in comparison to ions and a value of $\delta_{i}=5$
indicates $80\%$ depletion. We note that in the ion-acoustic time
scale, the dust particles does not contribute to the wall potential
but they modify the potential in the sheath by just being there as
a huge collections of negative charges, requiring more number of ions
to neutralise in the sheath away from the wall as compared to a sheath,
devoid of dust. As a result, the sheath must get thicker as we move
away from the wall. This effect can be seen in Fig.\ref{fig:potentials-1}.

\section{Forces on dust particles}

There may be several forces which act simultaneously on a dust particle\cite{13}.
The most significant of these are electrostatic, gravity, polarisation,
ion drag, and neutral drag force. The total force $\bm{F}$ on the
dust particles (dimensional) can be written as,
\begin{equation}
\bm{F}=\bm{F}_{E}+\bm{F}_{{\rm pol}}+\bm{F}_{g}+\bm{F}_{{\rm ion}}+\bm{F}_{{\rm n}},
\end{equation}
where the forces are respectively electrostatic, polarisation, gravity,
ion drag, and neutral drag force. The electrostatic force is given
by $\bm{F}=Q_{d}\bm{E}$. The polarisation force $\bm{F}_{{\rm pol}}=\nabla(\bm{{\cal P}}\cdot\bm{E})$
can be important when shielding cloud around a dust particle becomes
distorted due to polarisation, $\bm{{\cal P}}$ being the dipole moment.
The polarisation force can be neglected except for very dense dusty
plasmas\cite{19} where there is a possibility of distortion of the
charged cloud around a dust particle. The force due to gravity is
given by $m_{d}\bm{g}$, where $m_{d}$ is the mass of a dust particle
and $\bm{g}$ is the acceleration due to gravity. 

The ion drag force is given by\cite{20,21,22},
\begin{equation}
\bm{F}_{{\rm ion}}=\bm{F}_{{\rm coll}}+\bm{F}_{{\rm Coul}},
\end{equation}
where
\begin{equation}
\bm{F}_{{\rm coll}}=\pi r_{d}^{2}m_{i}n_{i}\bar{u}\bm{u}_{i}\left(1-\frac{2e\phi_{d}}{m_{i}\bar{u}^{2}}\right),\quad\bar{u}=\left(u_{i}^{2}+u_{s}^{2}\right)^{1/2},
\end{equation}
takes care of the momentum transfer due to ion-dust collision (collection
force). The Coulomb scattering part of ion-dust collision is given
by
\begin{equation}
\bm{F}_{{\rm Coul}}=4\pi b_{\perp}^{2}m_{i}n_{i}\bar{u}\bm{u}_{i}\ln\left(\frac{\lambda_{D}^{2}+b_{\perp}^{2}}{b_{{\rm max}}^{2}+b_{\perp}^{2}}\right)^{1/2}.
\end{equation}
In the above expression,
\begin{eqnarray}
b_{{\rm max}} & = & r_{d}\left(1-\frac{2e\phi_{d}}{m_{i}\bar{u}^{2}}\right)^{1/2},\\
b_{\perp} & = & \frac{eQ_{d}}{4\pi\epsilon_{0}m_{i}\bar{u}^{2}},
\end{eqnarray}
are the maximum impact parameter and impact parameter for $90^{\circ}$
scattering for an ion-dust collision.

The neutral drag force is given by\cite{23},
\begin{equation}
\bm{F}_{{\rm n}}=-\beta m_{d}\bm{u}_{d},\quad\beta=\delta\frac{8p}{\pi r_{d}m_{d}n_{d}u_{{\rm n}}},
\end{equation}
where $u_{{\rm n}}$ is the thermal velocity of the neutrals, $(\bm{u},m,n)_{d}$
are the dust velocity, mass, density, $p\equiv p_{i}$ is the gas
pressure, and $\beta$ plays the role of friction coefficient. The
parameter $\delta$ depends on details of the dust-neutral collision
and estimated as $\delta=1.26\pm0.13$\cite{24}.

\subsection{Simplified force model}

As we shall see that except the electrostatic force and force due
to gravity (assumed to be acting perpendicular to the sheath downward),
the other forces are negligibly smaller for the \emph{dust in plasma}
model considered in this paper and can be safely neglected (see below).

We note that in view of the ion-acoustic time scale, the dust particles
can be assumed to be stationary and we can assume that $u_{d}\ll u_{i}$
and so, the neutral drag force can be neglected. The remaining force
is the ion drag force and can be written as (normalized),
\begin{equation}
F_{{\rm ion}}=3R\bar{u}M\left[\frac{1}{4}r_{d}^{2}+b_{\perp}^{2}\ln\left(\frac{1+b_{\perp}^{2}}{b_{{\rm max}}^{2}+b_{\perp}^{2}}\right)^{1/2}\right],
\end{equation}
where the first term inside the `$[\,]$' is due to $F_{{\rm coll}}$
and the second one is due to $F_{{\rm Coul}}$ and the normalization
factor for $F$ is $(4/3)\pi\lambda_{D}^{3}\rho_{d}g$. The factor
$R$ is a dimensionless parameter representing the ratio of thermal
force on the electrons $(F_{p}=n_{0}T_{e}/\lambda_{D})$ to the gravitational
force on the dust particles $(F_{g}=\rho_{d}g)$
\begin{equation}
R=\frac{F_{p}}{F_{g}}=\frac{n_{0}T_{e}/\lambda_{D}}{\rho_{d}g},
\end{equation}
$\rho_{d}$ being the matter density of the dust particles. In the
above expressions, all parameters are normalized as before and the
normalized expressions for the impact parameters are\cite{20,21,22},
\begin{figure}
\begin{centering}
\includegraphics[width=0.5\textwidth]{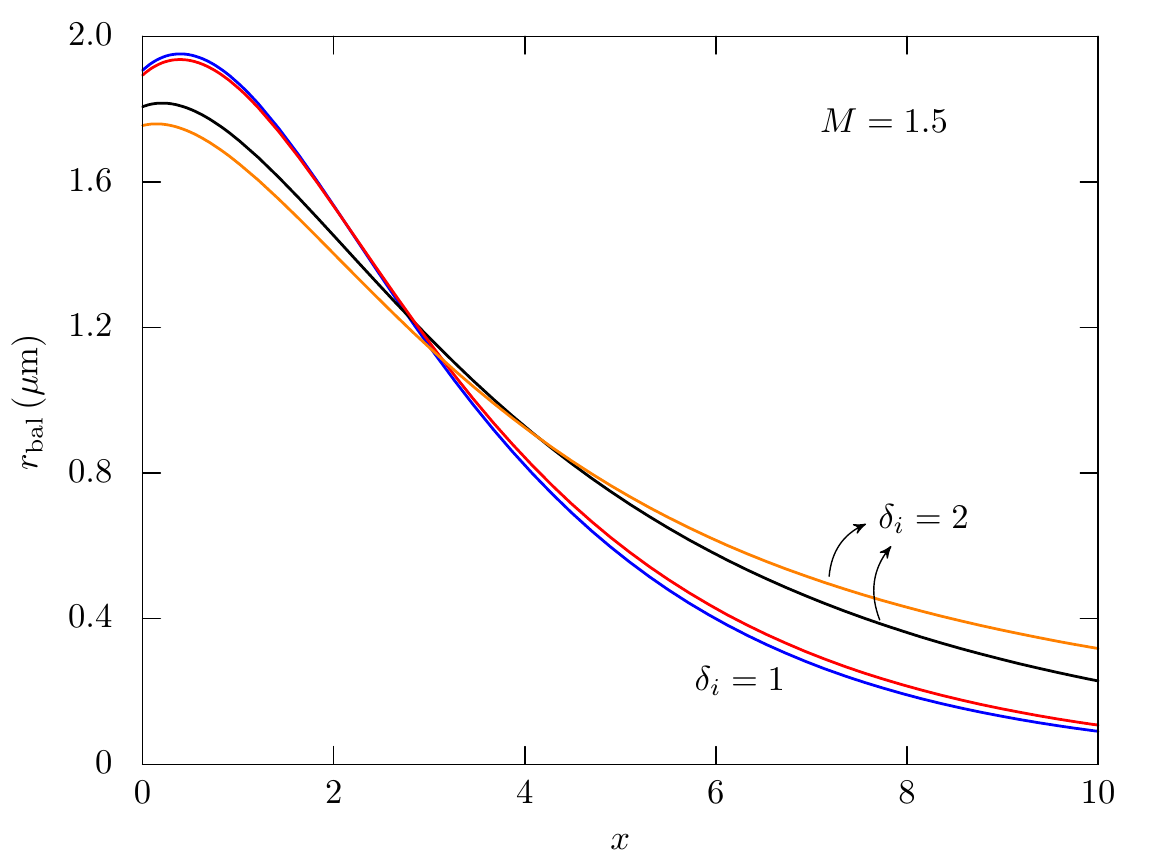}\hfill{}\includegraphics[width=0.5\textwidth]{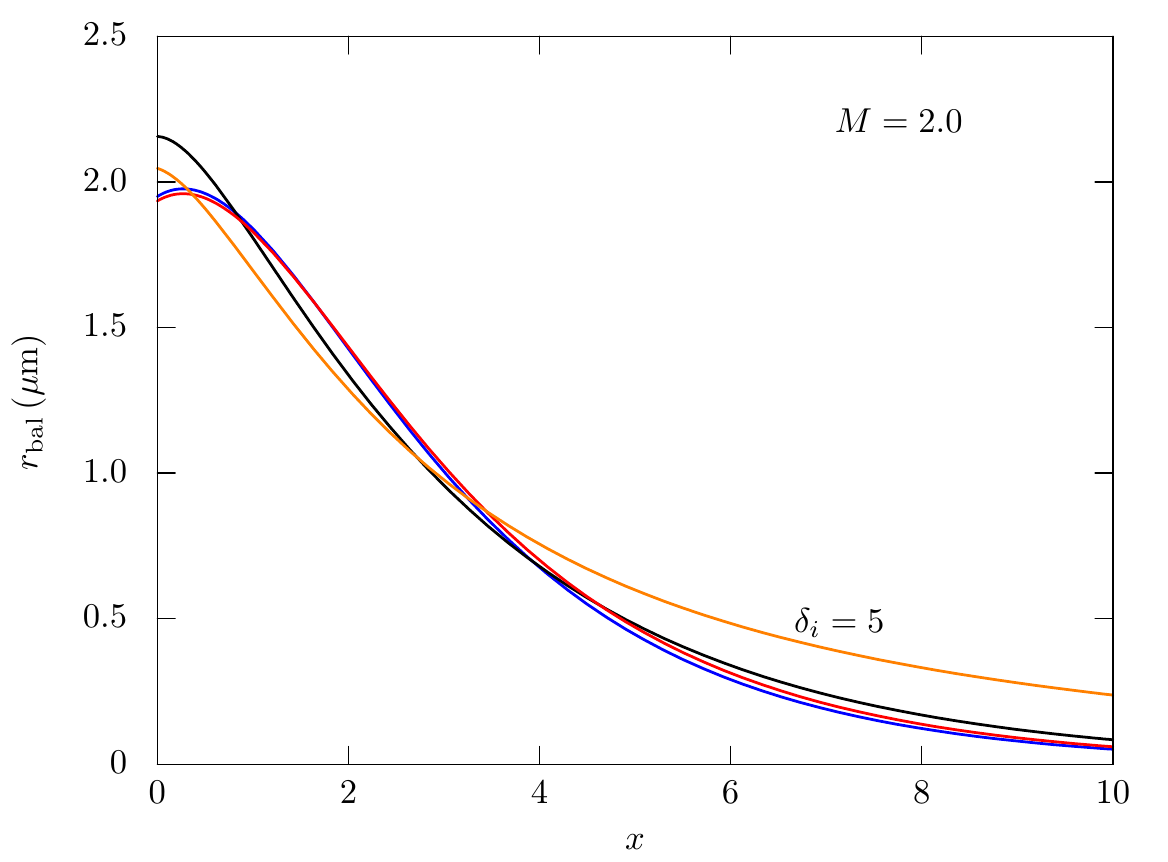}\\
\includegraphics[width=0.5\textwidth]{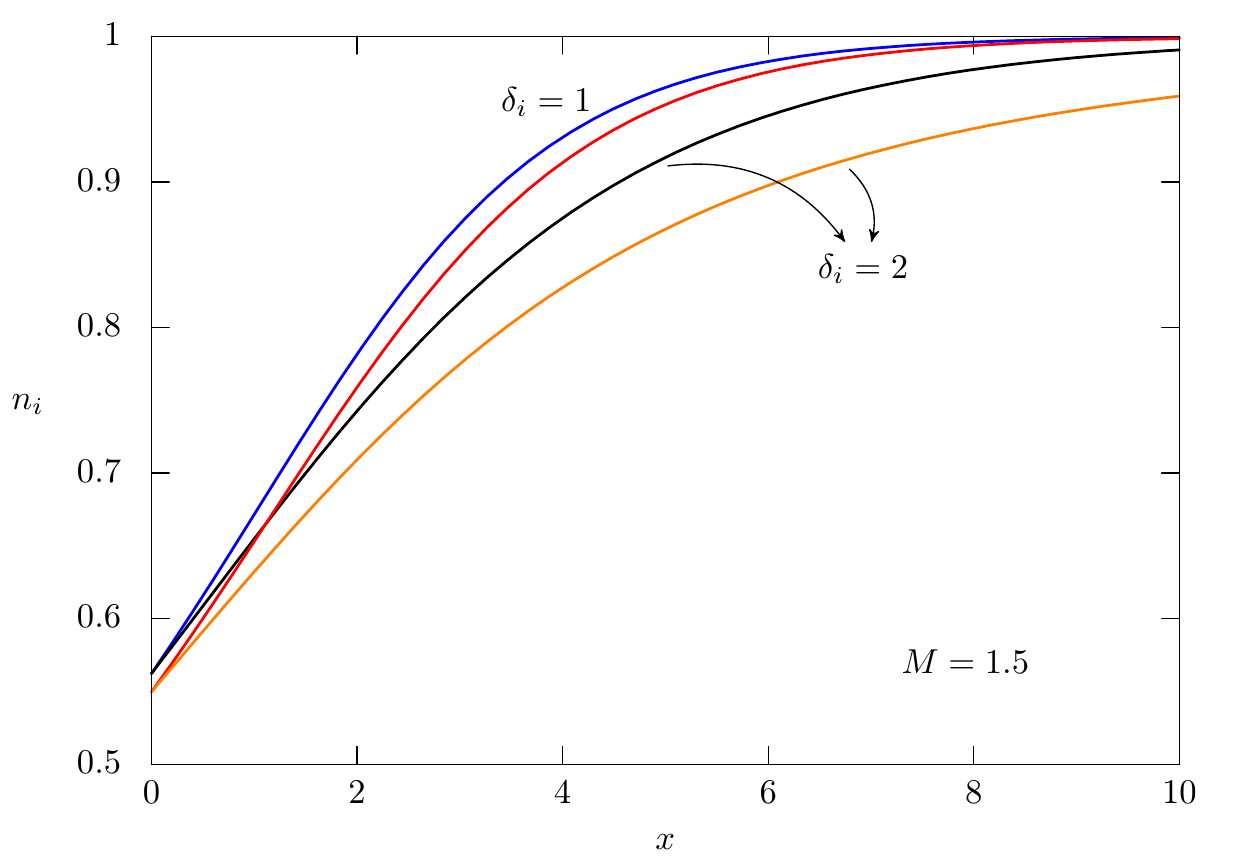}\hfill{}\includegraphics[width=0.5\textwidth]{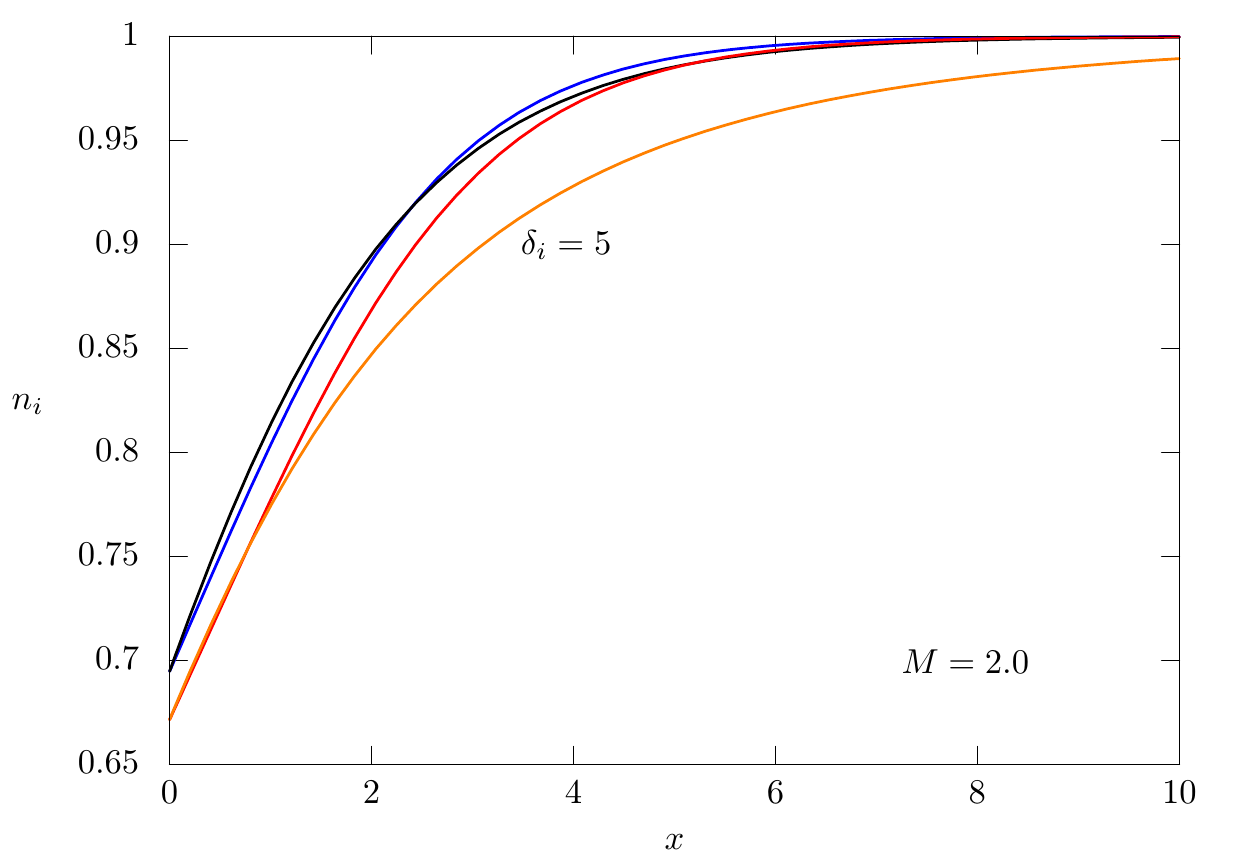}
\par\end{centering}

\protect\caption{\label{fig:balance}Distribution of size of levitated dust particles
$(r_{{\rm bal}})$ (top panel) and ion density (bottom panel) with
normalized distance $(x)$ from the wall in the sheath region. The
parameters and color-codes are same as in Fig.\ref{fig:potentials-1}.}
\end{figure}
\begin{eqnarray}
b_{{\rm max}} & = & r_{d}\left(1-\frac{2\phi_{d}}{\bar{u}^{2}}\right),\\
b_{\perp} & = & r_{d}\frac{\phi_{d}}{\bar{u}^{2}},
\end{eqnarray}
where
\begin{equation}
\bar{u}=\left(\frac{M^{2}}{n_{i}^{2}}+1\right)^{1/2}.
\end{equation}
With $F_{E,g,{\rm ion}}$, the normalized expression for total force
can be written as,
\[
F=-3r_{d}R\phi_{d}\left(\frac{d\phi}{dx}\right)-r_{d}^{3}+3R\bar{u}M\left[\frac{1}{4}r_{d}^{2}+b_{\perp}^{2}\ln\left(\frac{1+b_{\perp}^{2}}{b_{{\rm max}}^{2}+b_{\perp}^{2}}\right)^{1/2}\right],
\]
which is a highly nonlinear function of $r_{d}$. If we now assume
that $\bm{F}_{{\rm ion}}$ is negligible, the total force perpendicular
to the plasma sheath can be written as,
\begin{equation}
F=r_{d}\left(r_{{\rm bal}}^{2}-r_{d}^{2}\right),\label{eq:total_force}
\end{equation}
where
\begin{equation}
r_{{\rm bal}}=\left[-3R\phi_{d}\left(\frac{d\phi}{d\xi}\right)\right]^{1/2}.
\end{equation}
The quantity $r_{{\rm bal}}$ indicates the normalized radius of the
dust particles for which the total force on the dust particles become
zero, which leads to levitation of the dust particles in the plasma
sheath. A plot of $r_{{\rm bal}}$ is shown in Fig.\ref{fig:balance}.
As a prototype case, we have considered plasma sheath over lunar surface
and have taken various parameters as $n_{i0}=5\times10^{6}\,{\rm m}^{-3},\rho_{d}=1000\,{\rm kg/{\rm m}^{3}},T_{e}=50\,{\rm eV},g=1.6\,{\rm m}/{\rm s}^{2}$\cite{6}.
If we now calculate the magnitudes of different forces on a dust particle
for these parameters assuming an average radius for dust particle
to be $\sim10^{-6}\,\mu{\rm m}$, at $x=0$, the forces (normalized)
are $F_{E}\sim2.8\times10^{-22},F_{g}\sim-7.9\times10^{-23}$, and
$F_{{\rm ion}}\sim2.2\times10^{-29}$ and we can see that 
\begin{figure}
\begin{centering}
\includegraphics[width=0.5\textwidth]{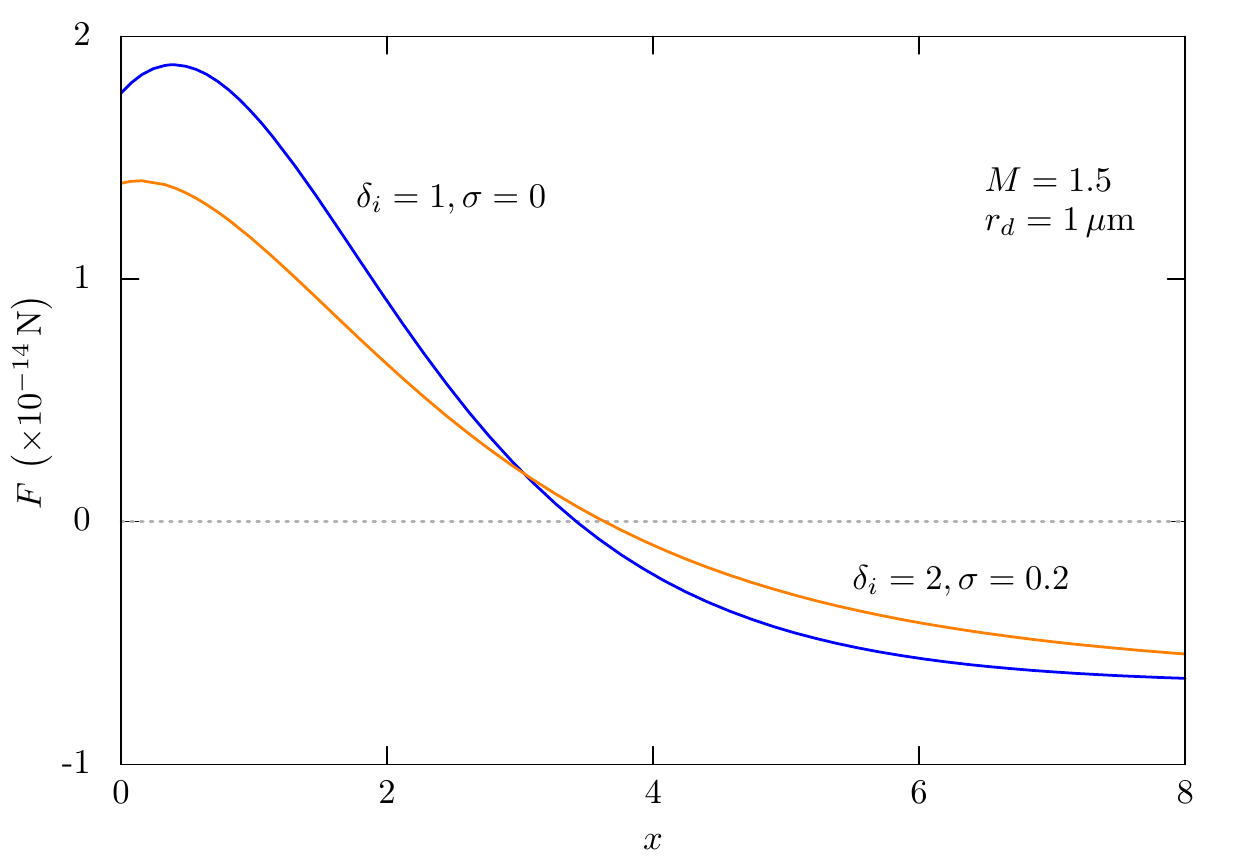}\hfill{}\includegraphics[width=0.5\textwidth]{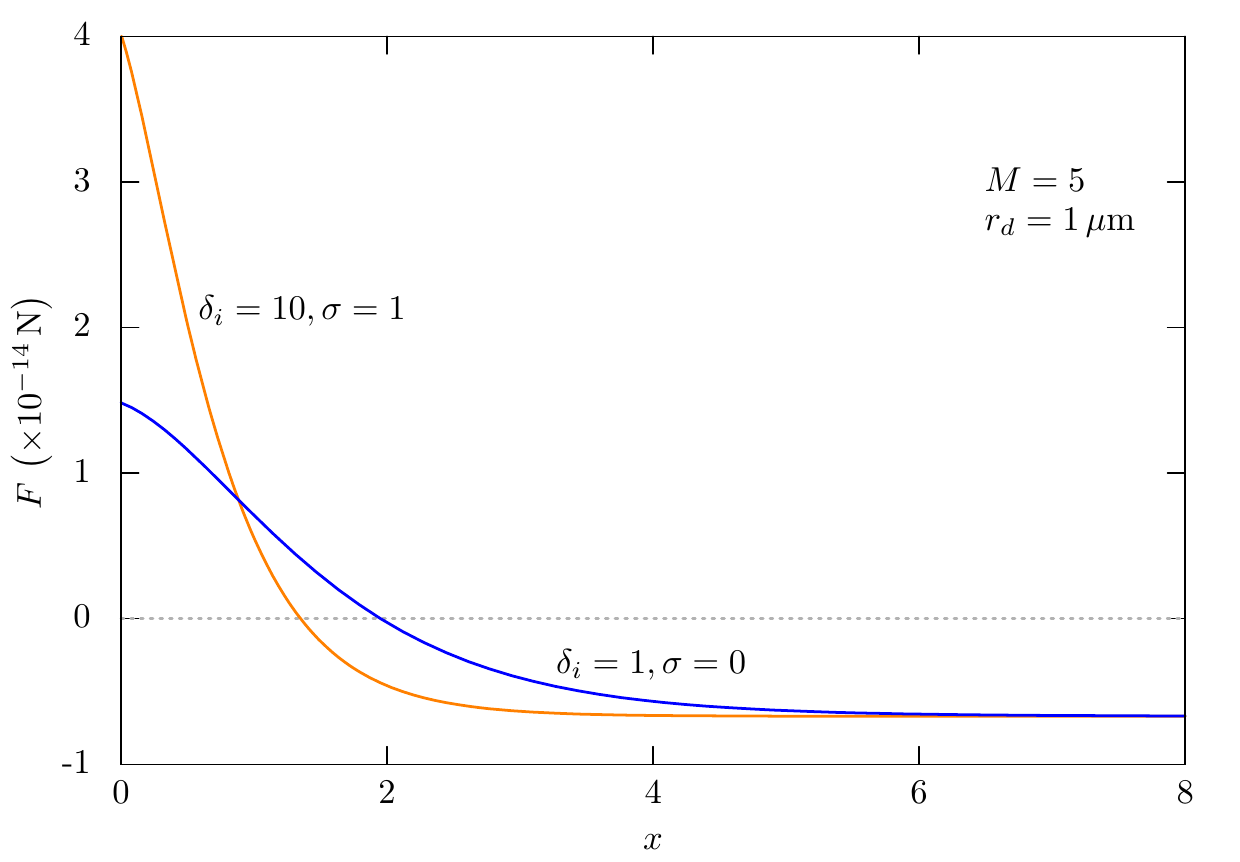}\\
\includegraphics[width=0.5\textwidth]{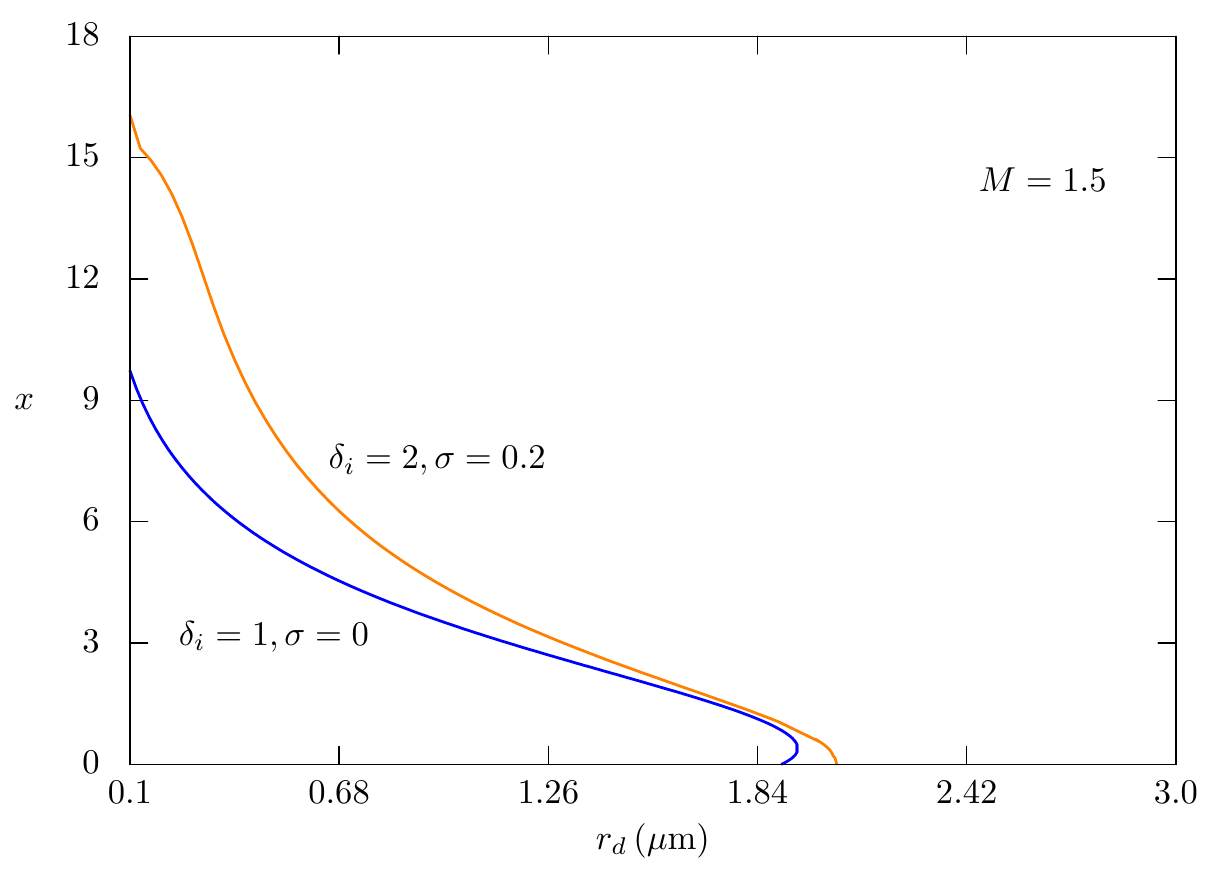}\hfill{}\includegraphics[width=0.5\textwidth]{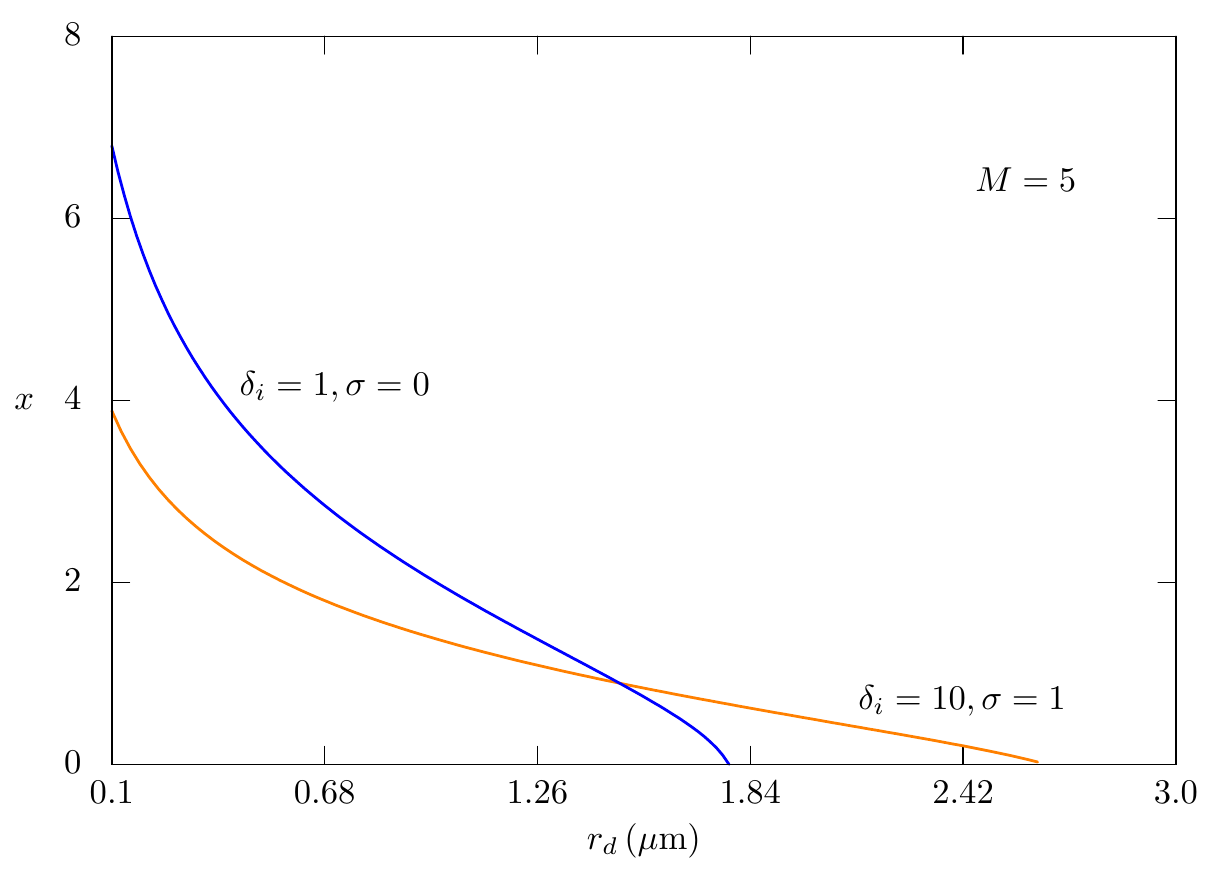}
\par\end{centering}

\protect\caption{\label{fig:F}The net force $F$ on the dust particles are shown in
the top panels for various parameters as indicated in the figure.
The corresponding locus of the point $F=0$ in the $x$-$r_{d}$ plane
is shown in the lower panels. Column-wise they correspond to same
sets of parameters. For details, see text.}
\end{figure}
\begin{equation}
F_{{\rm ion}}\ll F_{E,g},
\end{equation}
so that the total force on a dust particle can be approximated as
given by Eq.(\ref{eq:total_force}). We note that although we have
calculated the forces for a particular set of parameters, the above
condition remains valid for any values of parameters for this particular
model of \emph{dust in plasma}. As the size of a dust particle increases
it acquires more negative charges on its surface as well as also get
heavier. While more number charges (negative in this case) causes
the dust particle to repelled from the wall, the increasing weight
causes it to be attracted toward the wall, assuming the gravity to
be acting perpendicular to the wall downward. So, the exact position
where the dust particle remain levitated depends on it weight. In
this particular case, the gravity overtakes the electrostatic force
and larger dust particles of size $r_{d}\gtrsim1.75\,\mu{\rm m}$
(for $\delta_{i}=2$) and $r_{d}\gtrsim2\,\mu{\rm m}$ (for $\delta_{i}=5$)
always settles down on the wall (see Fig.\ref{fig:balance}). At the
same time, as dust particles gets more laden with negative charges,
the increasing electrostatic force expels the dust particle away from
the wall and the levitation distance from the wall increases. As can
be seen from Fig.\ref{fig:balance}, the levitation distance from
the wall increases from $2.8\lambda_{D}$ to $3.6\lambda_{D}$ by
about 30\% for a dust particle of $1\,\mu{\rm m}$ size when $\delta_{i}$
increases to $5$ from $2$. The ion density also follows the dust
distribution.

The net force $F$ on a dust particle is shown in Fig.\ref{fig:F}
(top panel). All parameters are indicated in the figure. As can be
seen, the force passes through the zero axis indicating the position
in the sheath where a particular dust particle $(r_{d}=1\,\mu{\rm m})$
levitates. The Mach numbers are $1.5$ and $5$. The top panel on
the left is for $\sigma=0$ and $0.2$ and on the right is for $\sigma=1$
(equal ion and electron temperature). The blue curves on both panels
are for $\delta_{i}=1$ and the orange curves are for $\delta_{i}=10$
signifying $90\%$ depletion of electron on the surface of dust particles.
The lower panels correspond to the respective parameters of the top
panels but show the locus of the point where the net force $F=0$
in the $x$-$r_{d}$ plane. From this figure, the effect of higher
dust concentration in the plasma sheath can be clearly seen. For both
low and high Mach numbers, the behaviour of $F$ with $x$, remains
almost same when the dust effect is not considered. For lower Mach
numbers (left panes), hot ions $(\sigma=0.2)$ causes the sheath to
expand which pushes the dust particles away from the wall which is
well understood. However for high Mach numbers (right panels), increase
of dust particles inside the sheath prevents this push and they tend
to settle down closer to the wall, even when the ions are now as hot
as the electrons $(\sigma=1)$. 
\begin{figure}
\begin{centering}
\includegraphics[width=0.5\textwidth]{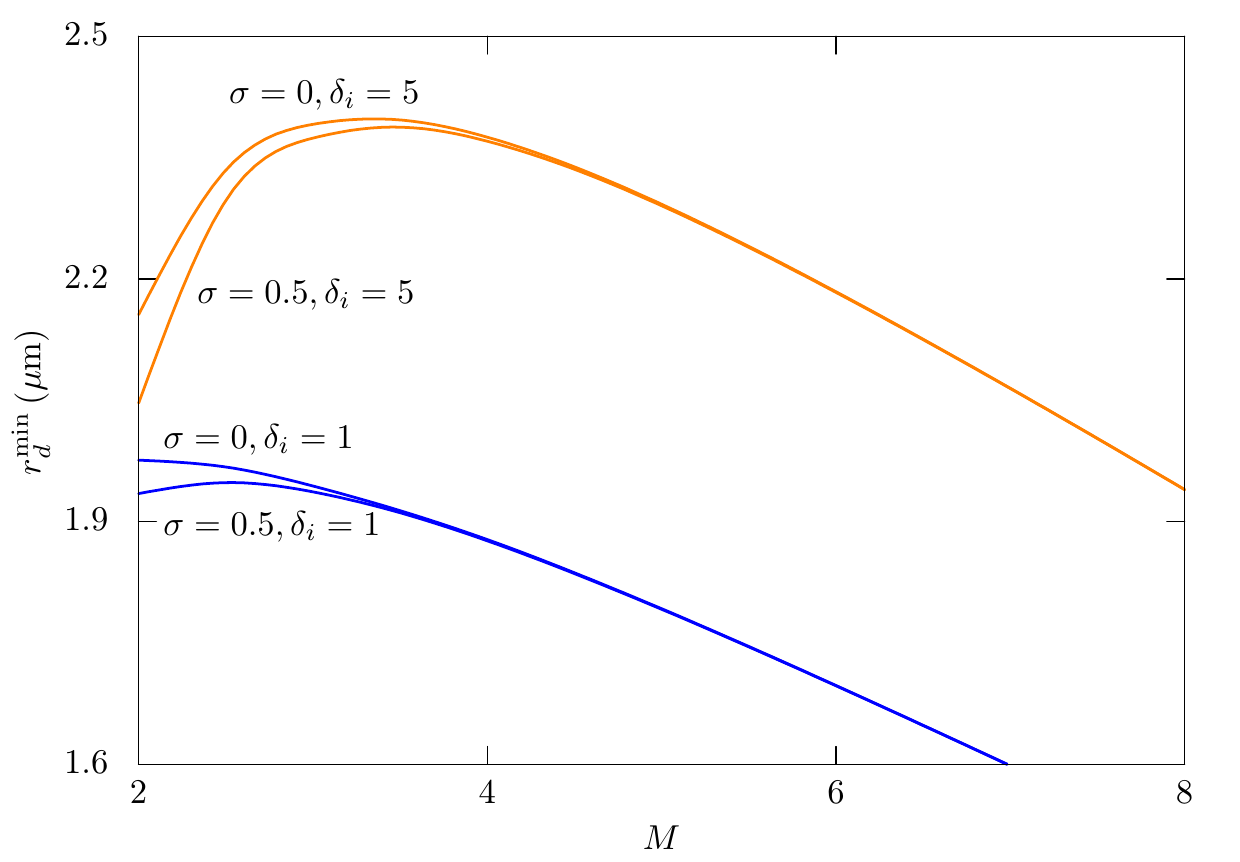}
\par\end{centering}

\protect\caption{\label{fig:rmin}Variation of $r_{d}^{{\rm min}}$ with $M$ for cold
and warm sheath with and without dust effects. All parameters are
same as in earlier figures.}
\end{figure}

Levitation of dust particles in the sheath is due to the balance of
the electrostatic force and the force due to gravity. Heavier dust
particles always settle down on the wall irrespective of the repulsive
electrostatic force on them which gives rise to a minimum size $r_{d}^{{\rm min}}$,
for the dust particles for levitation. This can be calculated by following
the locus of the maxima of the the net force $F(x)$ as a function
of $x$ in the $x$-$r_{d}$ plane. From Eq.(\ref{eq:total_force}),
the maximum of $F(x)$ is given by,
\begin{equation}
\frac{d}{dx}r_{{\rm bal}}=0,
\end{equation}
which translates to the condition,
\begin{equation}
\frac{d}{dx}\left(\phi_{d}\frac{d\phi}{dx}\right)=\phi_{d}'\phi'+\phi_{d}\phi''=0,
\end{equation}
where the prime denotes derivative with respect to $x$. The above
equation can be solved for $x$ resulting $x_{{\rm max}}$ such that
$F'(x_{{\rm max}})=0$. We then need to solve the equation
\begin{equation}
F(r_{d})|_{x_{{\rm max}}}=0
\end{equation}
for $r_{d}$, which determines $r_{d}^{{\rm min}}$. In Fig.\ref{fig:rmin},
we have shown the numerically calculated $r_{d}^{{\rm min}}$ as a
function of the Mach number. Though more number of dust particles
in the sheath restricts the size of dust particles nearer the wall,
it nevertheless pushes up the minimum size of the dust particles,
which can be levitated in the sheath. As a result, due to the dust
effecting the plasma potential inside the sheath region, we can expect
a significant reduction of the distance away from the wall up to which
we have dust levitation, but the levitated dust particles should now
have an extended distribution of size as $r_{d}^{{\rm min}}$ increases.

\section{Conclusions}

In this work, we have considered the formation of warm plasma sheath
in the vicinity of a wall in an environment with a considerable presence
of dust particles. As an example, we have used the parameters relevant
in case of lunar plasma sheath, though the results obtained in this
work could be applied to any other physical situation such as laboratory
plasma. In the ion-acoustic time scale, we have neglected the dust
dynamics and thus the dust particle provides a stationary background
to the electron-ion plasma. However, the dust particles do affect
the sheath dynamics by affecting the Poisson equation which determines
the plasma potential in the sheath region. We have assumed that the
current to a dust particle remains balanced throughout the sheath
formation process. This makes the grain potential dependent on plasma
potential, which is then incorporated into the Poisson equation. The
resultant numerical model becomes an initial value problem, which
is described by a 1-D integro-differential equation, which is then
solved self-consistently by incorporating the change in plasma potential
caused by inclusion of the dust potential in the Poisson equation.
The initial value is given by the plasma potential at the wall determined
by the Mach number (equilibrium ion thermal velocity).

We have shown that the presence of massive dust particles inside the
sheath region considerably affects the plasma sheath. As dust particles
are introduced into the sheath, which may happen naturally in case
of space plasma environments and through plasma-wall interactions
in laboratory plasmas, they becomes sources of huge collection of
negative charges (in this case) requiring more number of ions to neutralise
the negative charges while the wall potential remain same. As a result,
the sheath becomes thicker. The warm ions helps this process of thickening
much like a Debye shielding expanding due to temperature.

We have also considered the phenomenon of levitation of dust particles
inside the plasma sheath, which happens due to counteracting electrostatic
force and force due to gravity (assuming gravity to be acting perpendicularly
downward tom the sheath). Though there may be many other sources of
force on a dust particle, they can be neglected in most cases. Bigger
dust particles acquire more charges on their surfaces though at the
same time get heavier. As a result they may get levitated inside the
sheath although the levitation distance (as measured in terms of distance
from the wall) may differ depending on the constituent of dust particles.
For low Mach numbers, hot ions causes the sheath to expand which pushes
the dust particles away from the wall. This effect is like the expansion
of Debye shielding cloud expanding due to temperature. For high Mach
numbers, however, increase of dust particles inside the sheath prevents
this push and they tend to settle down closer to the wall, even when
the ions are as hot as electrons. Heavier dust particles always settle
down on the wall, which gives rise to a minimum dust size which is
required for levitation. Increasing dust density in the plasma sheath
restricts the size of levitating dust particles nearer the wall. However,
on the average, it pushes up the minimum levitation size. As a result,
while due to the effect of dust potential affecting the floating plasma
potential in the sheath causes a significant reduction of the levitation
distance, the dust distribution becomes broader.

In summary, our analysis provides a generic framework for dust in
plasma sheath which could be paramerized to suit laboratory experiments
as well as different space and astrophysical environments. To analyse
the detailed distribution of levitating dust particles in a plasma
sheath in a particular physical scenario, we need to input the respective
physical parameters into our analysis. The levitating dust particles
in plasma sheath may very well affect results of various plasma processing
experiments and the knowledge of detailed dust distribution in plasma
sheath is desired to tailor particular needs of experiments. 
\begin{acknowledgments}
One of the authors, R.D.\ would like to thank UGC for financial grant
provided through RFSMS Fellowship.\end{acknowledgments}

\end{document}